\documentclass[fleqn,usenatbib]{mnras}

\usepackage{newtxtext,newtxmath}
\usepackage[T1]{fontenc}
\usepackage{graphicx}	
\usepackage{amsmath}	
\usepackage{xspace}
\usepackage{booktabs}
\usepackage{hyperref}
\usepackage{upgreek}
\usepackage[dvipsnames]{xcolor}

\title[f(R) emulator]{FORGE --- the $f(R)$ gravity cosmic emulator project I: Introduction and matter power spectrum emulator}

\author[C. Arnold et al.]{
Christian Arnold,$^{1}$\thanks{E-mail: christian.arnold@durham.ac.uk}
Baojiu Li$,^{1}$\thanks{E-mail: baojiu.li@durham.ac.uk}
Benjamin Giblin,$^{2}$ 
Joachim Harnois-Déraps,$^{2,3,4}$ \newauthor
and Yan-Chuan Cai$^{2}$ 
\\
$^{1}$Institute for Computational Cosmology, Department of Physics, Durham University, South Road, DH1 3LE, Durham, UK\\
$^{2}$Scottish Universities Physics Alliance, Institute for Astronomy, University of Edinburgh, Blackford Hill, Scotland, UK\\
$^3$School of Mathematics, Statistics and Physics, Newcastle University, Herschel Building, NE1 7RU, Newcastle-upon-Tyne, UK\\
$^4$Astrophysics Research Institute, Liverpool John Moores University, 146 Brownlow Hill, Liverpool, L3 5RF, UK\\
}

\newcommand{\fr}{$f(R)$-gravity\xspace}
\newcommand{\fR}{\ensuremath{f_R}}
\newcommand{\fz}{\ensuremath{\bar{f}_{\mathrm{R0}}}}
\newcommand{\lcdm}{$\Lambda$CDM\xspace}
\newcommand{\mpcoh}{\ensuremath{\, h^{-1}\mathrm{Mpc}}}
\newcommand{\gpcoh}{\ensuremath{\, h^{-1}\mathrm{Gpc}}}
\newcommand{\hompc}{\ensuremath{\, h \mathrm{Mpc}^{-1}}}
\newcommand{\msoh}{\ensuremath{h^{-1} M_\odot}}

\begin{document}
\maketitle

\begin{abstract}
We present a large suite of cosmological simulations, the FORGE (F-of-R Gravity Emulator) simulation suite, which is designed to build accurate emulators for cosmological observables in galaxy clustering, weak gravitational lensing and galaxy clusters, for the $f(R)$ gravity model. The total of 200 simulations explore the cosmological parameter space around the \cite{planck2018} cosmology with a Latin hypercube, for 50 combinations of $\bar{f}_{\mathrm{R}0}$, $\Omega_\mathrm{m}$, $\sigma_8$ and $h$ with all other parameters fixed.
For each parameter combination, or node, we ran four independent simulations, one pair using $1024^3$ particles in $500\mpcoh$ simulation boxes to cover small scales, and another pair using $512^3$ simulation particles in $1.5\gpcoh$ boxes for larger scales. 
Each pair of initial conditions 
are selected such that sample variance on large scales is minimised on average. 
In this work we present an accurate emulator for the matter power spectrum in \fr trained on FORGE. 
We have verified, using the cross-validation technique, that the emulator accuracy is 
better than $2.5\%$ for the majority of nodes, particularly around the center of the explored parameter space, 
up to scales of $k = 10\hompc$. 
We have also 
checked the power spectrum emulator against simulations which are not part of our training set and found excellent agreement. Due to its high accuracy on small scales, the FORGE matter power spectrum emulator is 
well suited for weak lensing analysis 
and can play a key tool in constraining \fr using current and future observational data.
\end{abstract}

\begin{keywords}
cosmology: theory -- methods: numerical
\end{keywords}



\section{Introduction}

The origin of the accelerated Hubble expansion at late times---the cosmic acceleration---has been one of the most challenging questions in modern cosmology since its discovery over two decades ago \citep{SupernovaCosmologyProject:1998vns,SupernovaSearchTeam:1998fmf}. A huge body of research works have been conducted ever since, both in the development of theoretical models, such as dark energy \citep{Copeland:2006wr,Mortonson:2013zfa} and modified gravity \citep{clifton2012,joyce2015,Koyama:2020zce}, that can explain this phenomenon, and in observational efforts to test or constrain these models \citep[see, e.g.,][for general introductions of the various observational probes]{DETF2006,Weinberg2013}. While the standard $\Lambda$ cold dark matter (\lcdm) model, where the cosmic acceleration is assumed to be driven by a positive cosmological constant, $\Lambda$, has been widely accepted as the working model in the field, the profound implications of the discovery---that it could be a signature of new physics in either the particle or the gravity sector---have motivated extensive searches and investigations of other possibilities. Of these, modified gravity models have attracted a lot of attention in recent years, partly thanks to their potential to be used as testbeds to check the validity of General Relativity (GR) on length scales relevant for cosmology \citep[e.g.,][]{Koyama:2015vza,Ferreira:2019xrr,Baker:2019gxo}. 

Evidence of the cosmic acceleration has been supported by various independent cosmological observables/probes, in addition to the luminosity distance measurements with type Ia supernovae. These include the cosmic microwave background \citep[CMB; e.g.,][]{WMAP9, SPT2014,planck2018, ACT2020}, baryonic acoustic oscillations \citep[BAO; e.g.,][]{Cole2005_BAO,Eisenstein2005_BAO,Beutler2011_BAO,Blake2011_BAO,Anderson2012_BAO,eBOSS2020_BAO}, weak gravitational lensing \citep[e.g.,][]{Heymans2013,HSC_shear_corr_Y1, Asgari2021, Secco2021, Amon2021}, strong gravitational lensing \citep[e.g.,][]{Jullo2010_SL_constraints}, galaxy clusters \citep[e.g.,][]{Vikhlinin2009_Xray_cluster_count,Planck:2013lkt,Mantz:2014xba,Mantz:2014paa,SPT:2016izt,SPT:2018njh,DESY1_clusters,KiDS_clusters} and the clustering of galaxies \citep[e.g.,][]{Percival2004_2dF,Guzzo:2008ac,Beutler2012_growth_rate,Blake2011_wiggleZ_growth,BOSS:2016wmc,Pezzotta:2016gbo,Zarrouk:2018vwy}. In the coming years, several much larger galaxy surveys, such as DESI \citep{desi}, {\it Euclid} \citep{euclid} and the Vera C.~Rubin Observatory \citep{lsst} will be able to improve the current observational status by mapping the distribution of matter in the Universe with billions of galaxies. 

Given the significant improvement in statistical precision expected from these upcoming observatories, it becomes crucial that the accuracy on the prediction of many large-scale structure (LSS) observables reaches the percent level. A well-established approach to achieving this is by the use of cosmological $N$-body or hydrodynamical simulations, which are able to track the evolution of cosmic structures into the small-scale, highly nonlinear regime, where the multiple crossings of particle trajectories make it impossible to use linear perturbation theory to obtain reliable predictions. Although this approach is accurate and, with the rapid advancements of supercomputing power and generations of new simulation codes, has led to spectacular achievements in recent years \citep[e.g.,][]{Angulo2012_MXXL,schaye2015,Potter:2016ttn, pillepich2018, nelson2019}, they often come at a heavy computational cost and are therefore not well suited for sampling a large suite of models. More practical approaches often use lower resolution simulations, sparse cosmology sampling and approximations to the gravity solver \citep[see, e.g.,][]{Angulo2010_onesimfitall,Tassev:2013pn,Monaco:2013qta,Howlett2015,Feng:2016yqz,Klypin:2017iwu}. 

{The gravitational force in $N$-body simulations is computed from the Poisson equation, which can be solved using various algorithms such as `tree codes' as in \textsc{gadget} \citep{springel2005} and \textsc{arepo} \citep{springel2010}, multigrid relaxation as in \textsc{ramses} \citep[][]{Teyssier:2001_RAMSES_code_paper} or multigrid fast Fourier transform (FFT) such as in \textsc{cubep$^3$m} \citep{cubep3m}. However, in models where the theory of gravity differs from GR, the equation that describes gravity often becomes nonlinear in the matter field. This adds to the complexity of the problem and therefore accurate numerical solutions typically come at a significantly higher computational cost.} This makes the situation here comparatively worse than that described above for the $\Lambda$CDM model. There have been numerous codes developed to simulate the different modified gravity models studied in the literature \citep[e.g.,][]{oyaizu2008,schmidt2009c,Chan:2009ew,Li:2009sy,zhao2011b,li2012,puchwein2013,Li:2013nua,llinares2014,arnold2019,Hernandez-Aguayo:2020kgq}, thanks to which we have been able to gain qualitative and initial quantitative insights into the structure formation in these models. However, due to the impeded efficiency of modified gravity simulation codes, so far these simulations mostly suffer from two major limitations: the first is that their sizes (in terms of the box size, particle number and resolution) are significantly smaller than state-of-the-art simulations of \lcdm{} \citep[see, e.g.,][for several of the largest simulations of modified gravity models to date]{arnold2018}. The second is that the variety of models having been simulated is rather limited --- typically one fixes the cosmological parameters and varies only one (or a few of) modified gravity parameters to investigate the nonlinear effect on the large scale structure. However, not only are there potential degeneracies between modified gravity and certain cosmological parameters \citep[e.g.,][]{Baldi:2013iza}, but also the ability to predict cosmological observables when several---including cosmological---parameters are simultaneously varied is a key to constraining modified gravity models in the absence of prior knowledge of any of the parameters (especially when deviation from GR means that standard cosmological parameters can prefer values different from in \lcdm).

In this work, we will address the second limitation mentioned above, leaving the first one to be revisited in the future. {In particular, we consider the widely-studied $f(R)$ gravity \citep{buchdahl1970} model \citep[see, e.g.,][for recent reviews]{Sotiriou:2008rp,DeFelice:2010aj} with the $f(R)$ parameter $|\bar{f}_{R0}|$, and study its degeneracy with three cosmological parameters: the matter density $\Omega_{\rm m}$, the normalisation of the matter power spectrum $\sigma_8$, and the reduced Hubble parameter $h$. Ideally, one would sample a large number of points inside a wide parameter volume in order to constrain the model from observations, however the number of sampling points quickly explodes in a high-dimensional parameter space\footnote{{As an example, assuming that 10 points are sampled in each of the four dimension investigated here, then a total of $10^4$ models need to be simulated; even halving the number of points each dimension still results in $625$ models, which is well beyond the current available resources.}}. In fact it is already challenging to simulate $\mathcal{O}(100)$ $f(R)$ models with sufficiently large box size and high resolution, with the current generation of codes and supercomputers, making the grid approach not optimal.}

A useful technique to overcome this difficulty, which has been widely applied in cosmology research in recent years, is \textit{emulation}. With an optimal sampling of the model parameter space and interpolation of quantities from the sampled points, this makes it possible to make accurate predictions using a relatively small (e.g., $\lesssim\mathcal{O}(100)$) sample of points. The {cosmic emulation} technique \citep[e.g.,][]{Heitmann2006} was first applied to predict nonlinear matter power spectrum $P(k)$ with a $\%$-accuracy \citep{Heitmann2009,Heitmann2010,Heitmann2014}, where a {Latin hypercube} \citep{Heitmann2006} was used for the efficient sampling, ensuring the sampled points cover the whole parameter space as uniformly as possible. For effective interpolation, {Gaussian processes} \citep{gaussian_process}, a nonparametric Bayesian regression method, is most commonly used---though other approaches, such as neural networks \citep[e.g.,][]{Agarwal2014} are also employed.

Emulation has since been applied to predict various other physical and observable quantities, such as the galaxy correlation function \citep{Zhai2019}, halo mass function \citep{Mcclintock2019,Bocquet2020}, lensing shear correlation function \citep{slics}, or weak lensing peaks \citep{DESY1_peaks} and voids \citep{Davies2020} statistics. {These are enabled by dedicated suites of numerical simulations such as the \textsc{aemulus} \citep{aemulus}, {\it cosmo}-SLICS \citep{slics} and MassiveNuS \citep{MassiveNuS} projects.}

More recently, emulators have also been applied in the context of modified gravity models \citep[e.g.,][]{Winther2019, giblin2019, Ramachandra2020}, with the objective of providing accurate predictions of the matter power spectrum for Euclid \citep[][]{euclid} and the Vera Rubin Observatory \citep{lsst} respectively. This work can be considered as a continuation and extension of these previous works. While the variety of models we simulate and use to construct the $P(k)$ emulator is smaller than those used by \citet{Ramachandra2020}, our full $f(R)$ simulations are larger and have a higher resolution: {with} a total of 100 runs, for 50 cosmologies covering the above 4D parameter space, using a box of $500h^{-1}$Mpc and $1024^3$ particles, {our simulation program exploited}  $4$ million core-hours on the \textsc{cosma} machine, the UK's integrated supercomputing facility for theoretical modelling and HPC-based research in particle physics, astronomy and cosmology. The simulations were run using the modified gravity \textsc{arepo} code \citep{springel2010,arnold2019}, taking advantage of a substantial optimisation offered by the algorithm of \citet{bose2017}.

In this work, we introduce this suite of simulations, and present {a state-of-the-art $f(R)$ emulator for the fully nonlinear matter power spectrum, $P(k)$, that is trained on them. The matter power spectrum is one of the most powerful statistics since it captures all the statistical properties of an inhomogeneous, initially-Gaussian, Gaussian density field, which encodes rich information about the structure formation history and its underlying physics. Furthermore, it can be directly related to other two-point statistics, such as galaxy correlation function and weak lensing power spectrum. These connections will be shown in upcoming papers based on alternative observables constructed from the same suite of simulations. In particular, for each model investigated in this paper, we have saved particle snapshots at pre-selected redshifts to construct past light-cones and therefore can enable the emulation of weak-lensing statistics in $f(R)$ and $\Lambda$CDM models.} We will present a similar emulator for another widely-studied class of models, the Dvali-Gabadadze-Porrati \citep[][(DGP)]{dvali2000} braneworld model, in a companion paper. 

This paper is organised as follows. In Section \ref{sect:theory} we briefly over the theory of $f(R)$ gravity and specify our particular model choice. In Section \ref{sect:sims} we describe the simulations designed to construct the matter power spectrum emulator in this work. Section will \ref{sect:emulator} present the results of the emulator, and finally we summarise and conclude in Section \ref{sect:con}.

\section{$f(R)$ gravity}
\label{sect:theory}

\fr \citep{buchdahl1970} is an extension of Einstein's general relativity (GR). It adds an additional degree of freedom to gravity, allowing it to vary in space and time. Within a GR framework one can define the action of \fr as
\begin{align}
S=\int {\rm d}^4x\, \sqrt{-g} \left[ \frac{R+f(R)}{16\pi G} +\mathcal{L}_{\rm m} \right],\label{action}
\end{align}
where $g$ is the determinant of the metric tensor $g_{\mu\nu}$ of the 4D (GR-like) manifold; $R$ is the Ricci scalar and $f(R)$ is a scalar function of $R$ which gives rise to the additional (scalar) degree of freedom; $G$ is the standard gravitational constant and $\mathcal{L}_{\rm m} $ denotes the matter Lagrangian. If one requires the variation of the action with respect to the metric to vanish, one can derive the field equations for (metric) \fr
\begin{align} 
G_{\mu\nu} + \fR R_{\mu\nu}-\left( \frac{f}{2}-\Box \fR\right) g_{\mu\nu} - \nabla_\mu \nabla_\nu \fR = 8\pi G T_{\mu\nu}, \label{Eequn}
\end{align} 
again in GR notation and {using Einstein's sum convention}, where $G_{\mu\nu}$ denotes the Einstein tensor, $R_{\mu\nu}$ is the Ricci tensor, $T_{\mu\nu}$ is the energy-momentum tensor and $\nabla_\mu$ the covariant derivative ($\box \equiv \nabla_\mu \nabla^\mu$). The quantity $f_R$ is the derivative of the scalar function with respect to the Ricci scalar $R$, $\fR \equiv \mathrm{d} f(R) / \mathrm{d} R$, and often called the scalar degree of freedom.

In the quasi static and weak-field limit (often refereed to as the Newtonian limit of \fr), the above equation simplifies to two equations: the modified Poisson equation,
\begin{align}
 \boldsymbol{\nabla}^2 \Phi = \frac{16\pi G}{3}\delta\rho - \frac{1}{6} \delta R,\label{poisson}
\end{align}
and the equation of motion for the scalar degree of freedom $\fR$, that is obtained by taking the trace of Eq.~\eqref{Eequn},
\begin{align}
 \boldsymbol{\nabla}^2 \fR =  \frac{1}{3}\left(\delta R -8\pi G\delta\rho\right), \label{fRequn}
\end{align}
where $\delta \rho$ and $\delta R$ are respectively the perturbations to the density field and Ricci scalar, while $\boldsymbol{\nabla}$ denotes the gradient operator in three dimensions.

The modified Poisson equation shows that the total gravitational potential $\Phi$ and consequently the gravitational forces can be enhanced by a factor of $4/3$ in low curvature environments ($\delta R \approx 0$). If there was no mechanism to counterbalance this enhancement in higher density environments, this would of course lead to immediate tensions with Solar System constraints on gravity \citep{will2014}. On the other hand, \fr is well-known to feature the so-called chameleon screening mechanism \citep[][]{khoury2004,Mota:2006fz,brax2008}, which could drive $f_R\rightarrow0$ in regions with extensive deep Newtonian potential, therefore bringing the modified Poisson equation back to its behaviour in GR.

A model which is designed to comply with the local constraints on gravity is the one proposed by \citet{husa2007} using the following functional form for $f(R)$
\begin{align}
 f(R) = -m^2\frac{c_1\left(\frac{R}{m^2}\right)^n}{c_2\left(\frac{R}{m^2}\right)^n +1},\label{fr}
\end{align}
where $m^2 \equiv \Omega_\mathrm{m} H_0^2$, with $H_0$ being the Hubble constant. 
If one sets its free parameters, $c_1$ and $c_2$ to 
\begin{align}
\frac{c_1}{c2} = 6 \frac{\Omega_\Lambda}{\Omega_\mathrm{m}} && \text{and} && \frac{c_2\, R}{m^2} \gg 1,
\end{align}
the theory also features a \lcdm-like expansion history, fulfilling another important observational requirement \citep{husa2007}.  One can now simplify the differential equation for $\fR$ to
\begin{align}
\fR \equiv \frac{\mathrm{d} f(R)}{\mathrm{d} R} = -n\frac{c_1\left(\frac{R}{m^2}\right)^{n-1}}{\left[c_2\left(\frac{R}{m^2}\right)^n+1\right]^2}\approx-n\frac{c_1}{c_2^2}\left(\frac{m^2}{R}\right)^{n+1}.\label{fR}
\end{align}
For this paper, we will adopt $n=1$ as in most of the literature \citep[see, e.g.,][for studies of other values of $n$]{li2011,Ramachandra2020}. The remaining free parameters of the theory can now be re-expressed as $\bar{f}_{\mathrm{R0}}$, which is the background value of the scalar degree of freedom at $z=0$. It controls the potential depth threshold at which chameleon screening becomes active and GR-like forces are recovered. 

In our emulator simulations we shall vary $\fz$ between $\fz = -10^{-4.5}$ (F4.5) and $\fz=-10^{-6.2}$ (F6.2). F4.5 corresponds to a very strong modification of gravity which is in obvious conflict with current observational constraints \citep{terukina2014}. F6.2 on the other hand is a relatively weak modification of gravity where most high-mass haloes are completely screened. To make the emulator a useful tool for observational analysis, it is nevertheless necessary to cover this wide parameter range with our simulations. 

\section{The emulator simulation suite}
\label{sect:sims}

In this section, we describe the simulations used for constructing the $f(R)$ matter power spectrum emulators in this work. Our simulation suite consists of a total of 200 collisionless, dark-matter-only (DMO) runs covering 50 \fr models in a \lcdm background expansion. For each model, or `node', we ran two independent realisations with initial conditions designed to suppress the sampling variance \citep[as in][see Section \ref{sec:ics}]{slics}, for two different resolutions. The \textit{high-resolution} simulations were run with $1024^3$ DM particles in $500 \mpcoh$ side-length simulation boxes at a mass resolution of $m_\mathrm{part} = 9.1 \times 10^9 \msoh$. The \textit{low-resolution} simulations use $512^3$ DM particles in a $1500\mpcoh$ simulation box with a mass resolution of $m_\mathrm{part} = 1.5 \times 10^{12} \msoh$.
The gravitational softening length of the high-resolution runs is $15\, h^{-1}\mathrm{kpc}$ (the highest resolution of the AMR MG solver is similar) and the softening of the low-resolution simulations is $75\, h^{-1} \mathrm{kpc}$.

\subsection{Cosmological parameters}

\begin{figure}
	\includegraphics[width=\columnwidth]{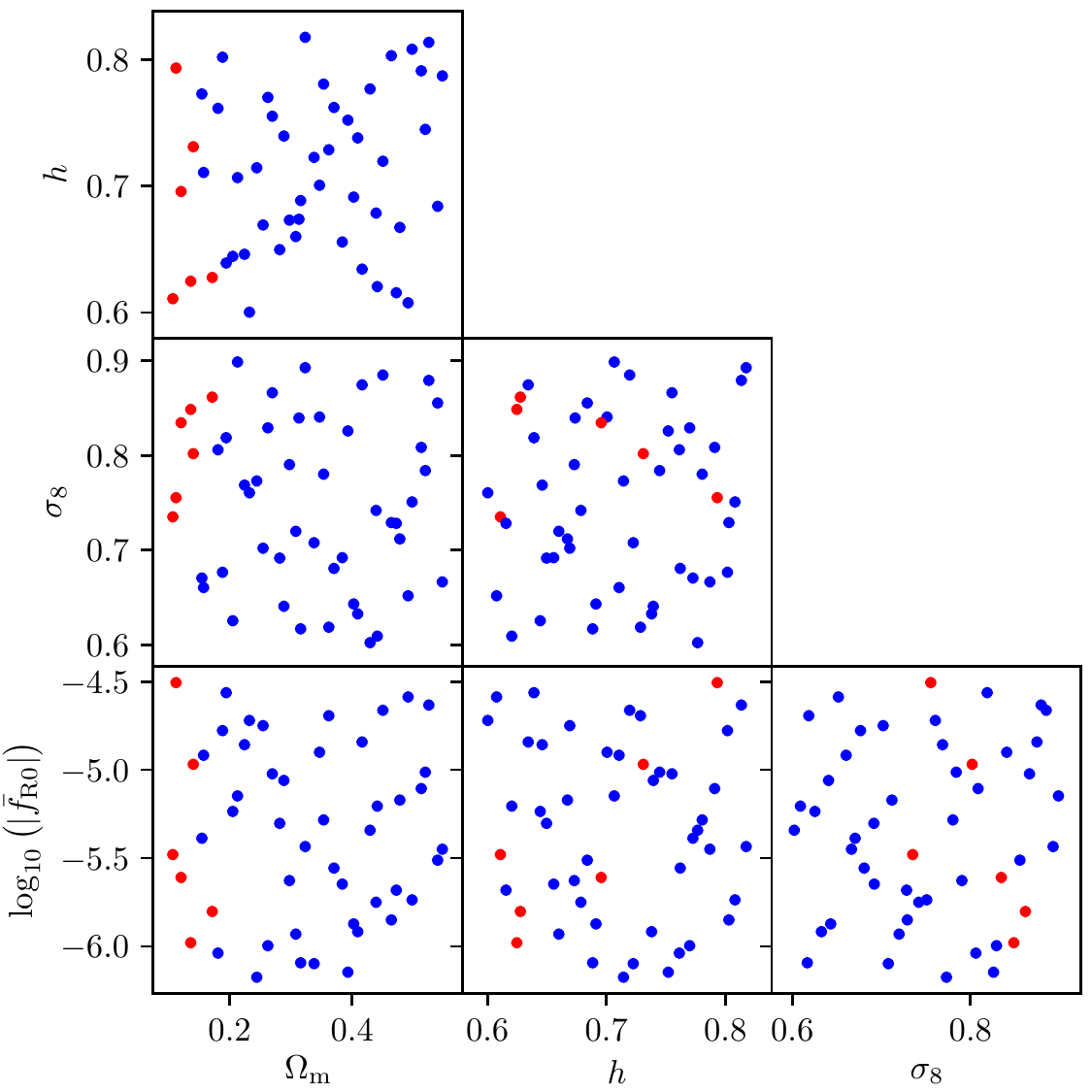}
    \caption{(Colour Online) Visualisation of the distribution of the parameters in the different sub-planes of the parameter-space studied. The fiducial cosmology simulation (node 0) is not shown in the planes involving $\log_{10}|\bar{f}_\mathrm{R0}|$ since $f_{R0}=0$ in that case. The nodes represented by red dots are the ones for which the emulator error exceeds $5\%$ at $z=0$ in a cross-validation test.}
    \label{fig:param}
\end{figure}

The simulations presented in this work explore the cosmological parameter space by varying $\Omega_\mathrm{m}$, $\sigma_8$, $h$ and $\bar{f}_\mathrm{R0}$ while keeping the other parameters fixed to $n_\mathrm{s} = 0.9652$, $\Omega_b = 0.049199$ and $\Omega_\Lambda = 1-\Omega_\mathrm{m}$. Because these simulations were partly designed to emulate weak lensing statistics,  we sample directly in the composite structure growth parameter $S_8\equiv\left(\Omega_{\rm m}/0.3\right)^{0.5}$
instead of the physical matter fluctuation amplitude parameter $\sigma_8$, and therefore better account for the degeneracy between $\Omega_{\rm m}$ and $\sigma_8$ in cosmic shear analyses.

The values for the variable parameters are listed in Table \ref{tab:param}. For our choice of parameters we take a similar approach as the \textsc{cosmo-slics} project \citep{slics}, sampling the parameters in a latin hyper-cube and ensuring that the parameters are as evenly distributed in all sub-planes of our parameter-space as possible. Due to the phenomenology of \fr, we do not sample evenly in $\bar{f}_\mathrm{R0}$ but in $\log_{10}|\bar{f}_\mathrm{R0}|$. The distribution of the variable parameters within the studied parameter-space is illustrated in Figure \ref{fig:param}.

\begin{table}
\centering
\begin{tabular}{l c c c c c c c}
\toprule
Node & $\Omega_{\rm m}$ & $\Omega_\Lambda$ & $\sigma_8$ & $h$ & $|\bar{f}_{\mathrm{R}0}|$\\
\midrule
0 & $0.31315$ & $0.68685$ & $0.82172$ & $0.67370$ & $0$ (\lcdm)\\
1 & $0.54725$ & $0.45275$ & $0.49342$ & $0.78699$ & $10^{-5.44975}$\\
2 & $0.53961$ & $0.46039$ & $0.63783$ & $0.68393$ & $10^{-5.51178}$\\
3 & $0.10721$ & $0.89279$ & $1.22974$ & $0.61090$ & $10^{-5.48008}$\\
4 & $0.31592$ & $0.68408$ & $0.60111$ & $0.68845$ & $10^{-6.09310}$\\
5 & $0.15741$ & $0.84259$ & $0.91175$ & $0.71067$ & $10^{-4.91748}$\\
6 & $0.35339$ & $0.64661$ & $0.71886$ & $0.78052$ & $10^{-5.28368}$\\
7 & $0.11240$ & $0.88760$ & $1.23413$ & $0.79318$ & $10^{-4.50605}$\\
8 & $0.39303$ & $0.60697$ & $0.72152$ & $0.75200$ & $10^{-6.14647}$\\
9 & $0.18096$ & $0.81904$ & $1.03776$ & $0.76132$ & $10^{-6.03817}$\\
10 & $0.42927$ & $0.57073$ & $0.50350$ & $0.77667$ & $10^{-5.34219}$\\
11 & $0.40249$ & $0.59751$ & $0.55523$ & $0.69120$ & $10^{-5.87285}$\\
12 & $0.21286$ & $0.78714$ & $1.06687$ & $0.70661$ & $10^{-5.14780}$\\
13 & $0.34671$ & $0.65329$ & $0.78191$ & $0.70056$ & $10^{-4.90056}$\\
14 & $0.15464$ & $0.84536$ & $0.93390$ & $0.77273$ & $10^{-5.38763}$\\
15 & $0.28172$ & $0.71828$ & $0.71367$ & $0.64968$ & $10^{-5.30326}$\\
16 & $0.37032$ & $0.62968$ & $0.61264$ & $0.76204$ & $10^{-5.55669}$\\
17 & $0.41627$ & $0.58373$ & $0.74242$ & $0.63427$ & $10^{-4.84239}$\\
18 & $0.32331$ & $0.67669$ & $0.85987$ & $0.81749$ & $10^{-5.43473}$\\
19 & $0.47784$ & $0.52216$ & $0.56403$ & $0.66724$ & $10^{-5.17131}$\\
20 & $0.20509$ & $0.79491$ & $0.75641$ & $0.64437$ & $10^{-5.23575}$\\
21 & $0.44103$ & $0.55897$ & $0.50237$ & $0.62046$ & $10^{-5.20564}$\\
22 & $0.46403$ & $0.53597$ & $0.58620$ & $0.80296$ & $10^{-5.85013}$\\
23 & $0.13644$ & $0.86356$ & $1.25837$ & $0.62473$ & $10^{-5.97960}$\\
24 & $0.18832$ & $0.81168$ & $0.85396$ & $0.80174$ & $10^{-4.77781}$\\
25 & $0.12066$ & $0.87934$ & $1.31591$ & $0.69563$ & $10^{-5.60979}$\\
26 & $0.28854$ & $0.71146$ & $0.65331$ & $0.73943$ & $10^{-5.06027}$\\
27 & $0.45016$ & $0.54984$ & $0.72241$ & $0.71954$ & $10^{-4.66274}$\\
28 & $0.17155$ & $0.82845$ & $1.13936$ & $0.62768$ & $10^{-5.80253}$\\
29 & $0.51949$ & $0.48051$ & $0.59577$ & $0.74473$ & $10^{-5.01340}$\\
30 & $0.43909$ & $0.56091$ & $0.61327$ & $0.67856$ & $10^{-5.75021}$\\
31 & $0.49786$ & $0.50214$ & $0.58288$ & $0.80806$ & $10^{-5.73666}$\\
32 & $0.40909$ & $0.59091$ & $0.54179$ & $0.73799$ & $10^{-5.91685}$\\
33 & $0.23227$ & $0.76773$ & $0.86433$ & $0.60028$ & $10^{-4.72039}$\\
34 & $0.38390$ & $0.61610$ & $0.61174$ & $0.65570$ & $10^{-5.64729}$\\
35 & $0.26234$ & $0.73766$ & $0.88665$ & $0.76998$ & $10^{-5.99617}$\\
36 & $0.25453$ & $0.74547$ & $0.76212$ & $0.66918$ & $10^{-4.74984}$\\
37 & $0.29762$ & $0.70238$ & $0.79347$ & $0.67300$ & $10^{-5.62737}$\\
38 & $0.22423$ & $0.77577$ & $0.88911$ & $0.64603$ & $10^{-4.85759}$\\
39 & $0.30799$ & $0.69201$ & $0.71046$ & $0.66001$ & $10^{-5.93063}$\\
40 & $0.51288$ & $0.48712$ & $0.61834$ & $0.79098$ & $10^{-5.10624}$\\
41 & $0.14061$ & $0.85939$ & $1.17125$ & $0.73101$ & $10^{-4.96888}$\\
42 & $0.33782$ & $0.66218$ & $0.66702$ & $0.72256$ & $10^{-6.09796}$\\
43 & $0.52520$ & $0.47480$ & $0.66452$ & $0.81347$ & $10^{-4.63303}$\\
44 & $0.19435$ & $0.80565$ & $1.01717$ & $0.63911$ & $10^{-4.56309}$\\
45 & $0.26963$ & $0.73037$ & $0.91366$ & $0.75511$ & $10^{-5.02280}$\\
46 & $0.49135$ & $0.50865$ & $0.50927$ & $0.60766$ & $10^{-4.58728}$\\
47 & $0.47207$ & $0.52793$ & $0.58056$ & $0.61562$ & $10^{-5.68160}$\\
48 & $0.24424$ & $0.75576$ & $0.85676$ & $0.71436$ & $10^{-6.17488}$\\
49 & $0.36187$ & $0.63813$ & $0.56321$ & $0.72861$ & $10^{-4.69340}$\\
\bottomrule
\end{tabular}
\caption{The simulation parameters varied for the 50 nodes of the cosmic emulator simulations. An ASCII version of this table is available online: \url{https://bitbucket.org/arnoldcn/forge\_emulator/}}
\label{tab:param}
\end{table}

\begin{figure}
	\includegraphics[width=\columnwidth]{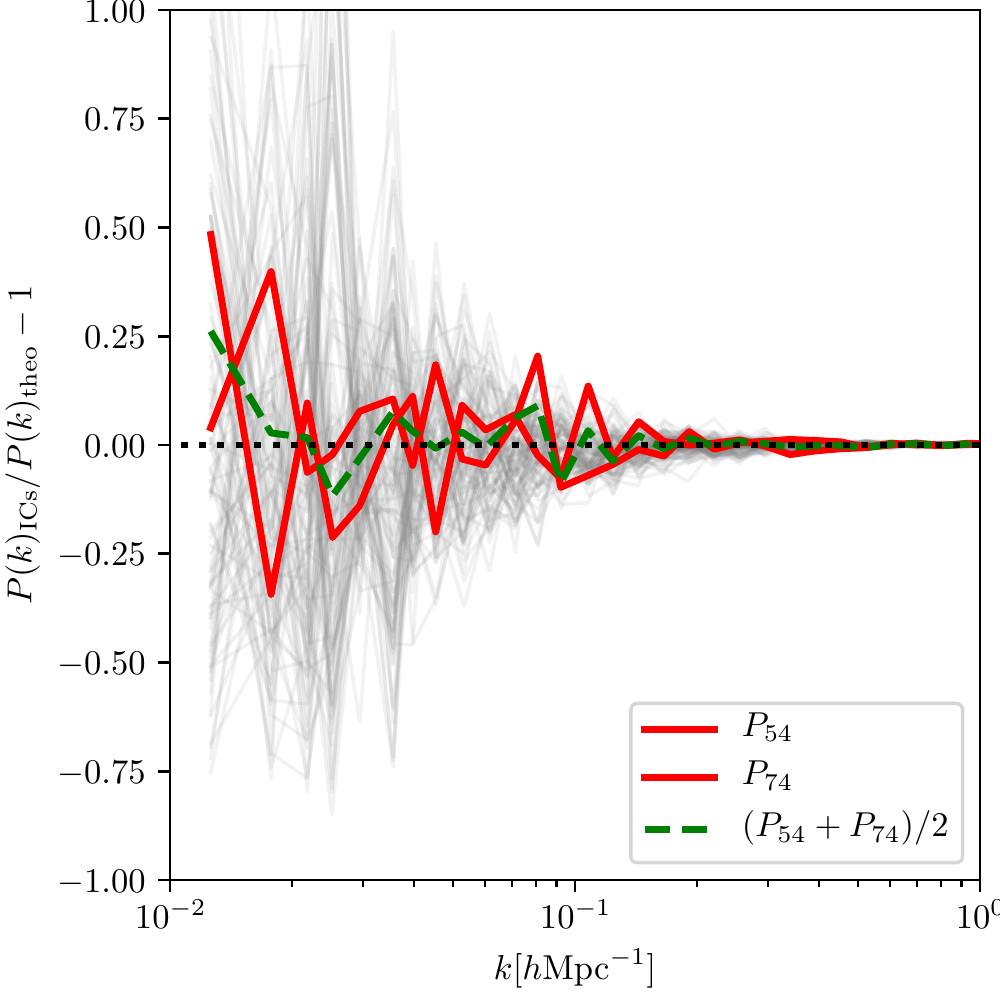}
    \caption{(Colour Online) The relative difference of the 100 independent IC power spectra to the theoretical input power spectrum (faint gray lines) as well as the $1\sigma$ and $2\sigma$ errors of the IC power spectrum distribution (blue shaded regions). The selected seeds are shown in red, and their mean by the green dashed line.}
    \label{fig:ic_power}
\end{figure}

\subsection{Initial condition generation}\label{sec:ics}

The power spectra of initial conditions (ICs) for cosmological simulations can never exactly resemble the theoretical input power spectrum which is used to generate the ICs at high redshift (in our case $z=127$). While the small scale differences due to the limited resolution of the ICs become irrelevant as the simulation progresses, the large scale errors or sample variance, which occur due to the limited box-size of the simulations are carried through while the simulations run and are still apparent at $z=0$. 

To limit these large scale errors, we run two implementations for a matched pair of ICs with independent phases (i.e., independent random seeds in \textsc{2LPT}) per node (cosmological parameter set) and box. The IC pair are selected such that their average large-scale error is as small as possible. In order to achieve this, we first generate $100$ independent ICs for node 0 (fiducial) cosmology and measure their power spectra. The relative difference of the actual IC power spectra to the theoretical input power spectrum is shown in Figure \ref{fig:ic_power}. From the 100 ICs, we selected two using the following criteria,
\begin{itemize}
\item the mean large scale error should be small,
\item their individual large scale error should be small,
\item and their error should cross the zero line as often as possible on large scales to avoid a leakage of power from large to small scales.
\end{itemize}
The two finally selected seeds for the ICs, 74 and 54, are shown in red in Figure \ref{fig:ic_power} and their mean as the green dashed line. We will refer to these implementations using their seeds, 2080 (74 in Fig. \ref{fig:ic_power})  and 4257 (54 in Fig. \ref{fig:ic_power}),  in the following.
{Notice that this technique does not produce a matched pair of ICs whose sample variance errors cancel exactly \citep[as in][]{angulo2016}, but instead follows \citet{slics} to generate a pair of ICs which (on average) roughly converges on the ensemble mean for the 3D power spectrum while each individual Fourier mode is still drawn from a Gaussian distribution. This is likely relevant for measurements of higher order statistics.}

The ICs were generated using the 2\textsc{lptic} \citep{lpt} code, an initial condition generator that employs second-order Lagrangian perturbation theory to compute particle displacements for a given initial matter power spectrum, based on \textsc{n-genic} \citep{springel2005}.

\subsection{MG simulations with \textsc{arepo}}

The simulations in this work were carried out with the \textsc{arepo} cosmological simulation code \citep{springel2010, weinberger2020}, using the modified gravity module presented in \citet{arnold2019}. In the following, we will only give a very brief overview of the code and refer the reader to the above references for more details. \textsc{arepo} uses a Tree Particle-Mesh algorithm to calculate the standard gravitational forces. The additional modified gravity force (fifth force) is calculated by employing a multigrid accelerated 
relaxation solver on an adaptive refining mesh (i.e., an AMR grid). 

To obtain the modified gravity forces, the code needs to solve Eq.~\eqref{fRequn} in the first place. In order to do so, the density field is binned onto the AMR grid, which is built such that on the finest refinement level each cell can contain at most one particle (except if a pre-set maximum refinement level is reached; the cell size at this level is of the order of the smoothing length of the standard gravity solver). Reformulating Eq.~\eqref{fRequn} as \citep[][]{bose2017}:
\begin{align}\label{eq:reduced_eom}
    \nabla^2(u^2) = \frac{1}{3} \left\{ \frac{\bar{R}(a)}{\bar{f}_\mathrm{R}(a)}  \left[ \frac{1}{u} - 1\right] - \frac{8\pi G}{\bar{f}_\mathrm{R}(a)}\delta\rho\right\}, 
\end{align}
where $u\equiv \sqrt{f_\mathrm{R} / \bar{f}_\mathrm{R}(a)}$ and over-bars denote background values, the solver now obtains the value of $f_\mathrm{R}$ iteratively on the grid. The MG force can then be easily computed from the gradient of the scalar field
\begin{align}
    \boldsymbol{a}_\mathrm{MG} = \frac{c^2}{2}{\boldsymbol{\nabla}}f_\mathrm{R},
\end{align}
where $c$ denotes the speed of light.

As the modified gravity solver can only compute the forces for all particles simultaneously, and the scalar field equation it solves is highly nonlinear, it is computationally more expensive than the standard gravity solver of \textsc{arepo}. The maximum modified gravity acceleration is however smaller than the maximum standard gravity acceleration, largely because 
the latter occurs in regions with high density (e.g., in massive haloes) that are screened from the fifth force contribution. This allows us to run the MG solver on larger time steps \citep[see][for details on the time-stepping scheme]{arnold2019}, significantly reducing the computational cost of the simulations.

Together with the efficient \textsc{mpi} parallelisation and lean memory footprint of \textsc{arepo}, these optimised numerical algorithms have made it possible to run the large number of $f(R)$ simulations used in this work. The 100 $500h^{-1}\textrm{Mpc}$ runs took roughly 3 million core hours on the \textsc{cosma}7 machine hosted at the Institute for Computational Cosmology, Durham University, by using 20 computer nodes, each with 512GB RAM and 28 cores (Intel Xeon Gold 5120 CPU @ 2.20GHz). The larger-box runs took much less time due to their lower resolution.

\section{An emulator for the 3D matter power spectrum}
\label{sect:emulator}

We use the simulations described above to build an emulator for the 3D matter power spectrum. Analyses of gravitational lensing properties, the halo mass function (HMF), lensing, etc., based on these simulations, will be presented in subsequent works. For the power spectrum emulator we rely on redshift outputs common to all nodes, namely at $z=0.0, 0.25, 0.50, 0.75, 1.00, 1.25, 1.50, 1.75$ and $2.00$. Below we present the measurement details and our emulator verification procedure, first focusing at $z=0$, however the measurements at larger $z$ are performed in the same way. A cross-validation of the emulator prediction at $z>0$ is shown in Appendix \ref{emu_ver}.

The power spectrum emulator is made publicly available online at \url{https://bitbucket.org/arnoldcn/forge\_emulator/}. A brief user guide can be found in Appendix \ref{emu_user_guide}.

\subsection{Power spectrum measurement}

The 3D matter power spectra are measured with the  estimator built into \textsc{arepo} employing the self-folding technique described in \citet{jenkins1998, springel2018}. This allows to obtain a 
high resolution for the power spectrum while avoiding computationally expensive FFTs on large grids. In this paper, we use an FFT resolution of $2048^3$ with a fold-factor $f_\mathrm{fold} = 4$ and apply the self folding twice (by using $f_\mathrm{fold} = 4^2$ for the second folding). This allows for a dynamic range of $32,000$ between the largest scales measured and the Nyquist frequency  of our  measurement. We apply this technique to both our small and our large box simulations, and both members of our matched IC pair. The resulting $3\times2\times2$ sections of the power spectra (3 foldings, 2 ICs, 2 resolutions) span different ranges in $k$ that overlap, allowing us to combine them into a single $P(k)$ per cosmological model. 

\begin{figure}
	\includegraphics[width=\columnwidth]{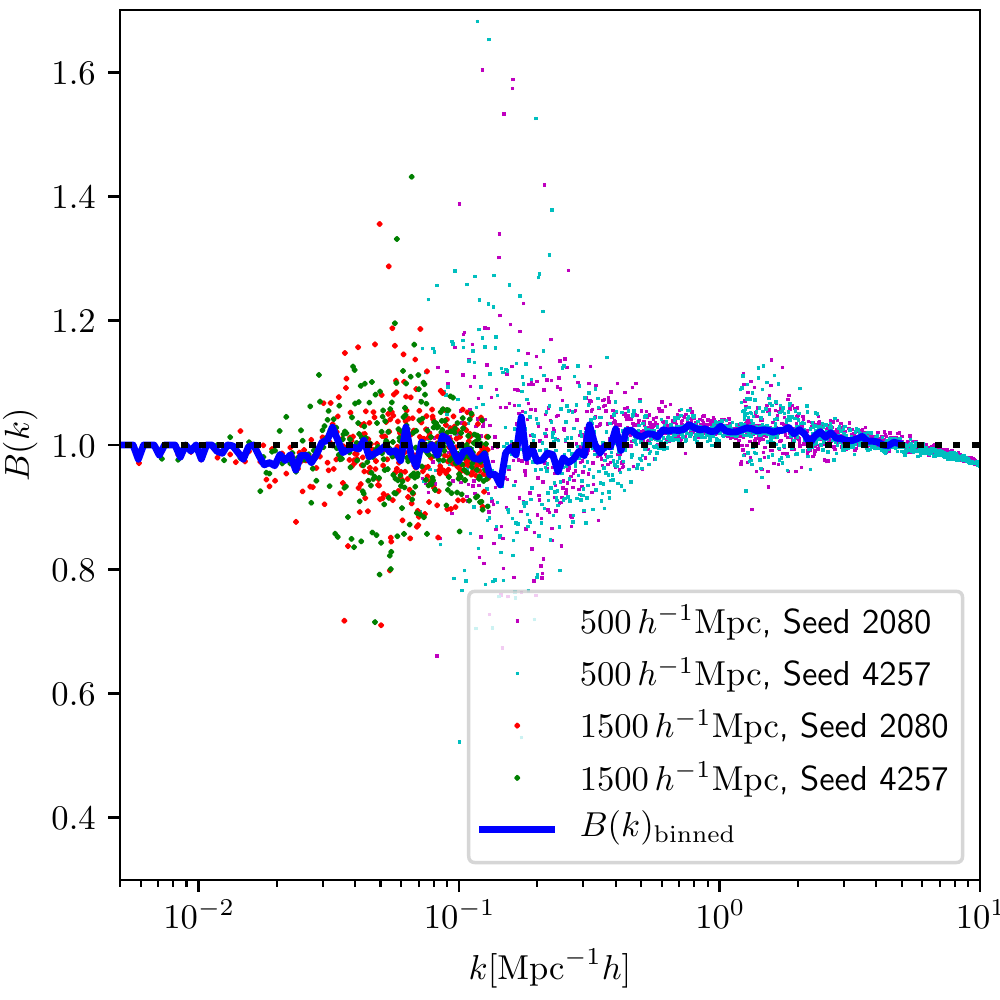}
    \caption{(Colour Online) The power spectrum response data ($B(k)$) for the different boxes and foldings of node 0, before (dots) and after (blue line) the binning. The black dotted line shows the equality to \textsc{halofit} prediction.}
    \label{fig:power_bin}
\end{figure}

To obtain a single power spectrum per cosmology, we divide the 12 sections by the 
\textsc{halofit} \citep{takahashi2012} nonlinear \lcdm prediction, to get the `response' defined as
\begin{equation}\label{eq:Bk_def}
    B(k) \equiv P(k)_\textrm{ simulation }^{ f(R) } \big/ P(k)_\textrm{ HALOFIT}^{\Lambda\textrm{CDM}}.
\end{equation} 
Note that the \textsc{halofit} prediction does not incorporate any \fr effects on the power spectrum, i.e., $B(k)$ will be a measure of the discrepancy between \textsc{halofit} and the simulation for our $\Lambda$CDM node (node 0), but will include modified gravity effects for all others. 
The aim of dividing by the theory prediction is to largely flatten out the $k$-dependence of the power spectrum. This way we can minimise the errors due to bin-centering and limit the dynamic range which has to be emulated, as described below. The 12 $B(k)$ sections (foldings) are finally binned into fine $k$-bins, from which our estimator extracts the median $B(k)$. Some sections of the large boxes are not actually used as they are dominated by shot noise. This binning process is illustrated  for node 0 in Figure \ref{fig:power_bin}. The different coloured symbols indicate the power spectra relative to \textsc{halofit} of the different implementations and boxes before binning, while the blue line after binning. Larger scatter appears towards the low-$k$ end of each folding.

The binned responses and power spectra for all nodes are shown in Figure \ref{fig:power}. To avoid an over-crowding plot, the data for the 50 nodes is split into 5 rows. The curves for the different nodes are colored according to the background field of the $f(R)$ model as shown in the colorbar at the top of the figure. Low numbers ($-\log_{10}\left(-\bar{f}_\mathrm{R0}\right) \approx 4$) mean 
less efficient chameleon screening 
and consequently stronger deviation from GR. Higher numbers ($-\log_{10}\left(-\bar{f}_\mathrm{R0}\right) \approx 6.5$) lead to stronger chameleon screening and a model behaviour more similar to GR.

The left column of Figure \ref{fig:power} shows the binned power spectra of the simulations (solid lines), as well as the linearly evolved initial power spectra\footnote{For linear evolution we have used the $\Lambda$CDM linear growth factor for each node with the cosmological parameters $\Omega_{\rm m}, h, \sigma_8$ for that node, assuming $\fR=0$.} (dotted lines). These panels also show the shot noise level of the simulations (the horizontal grey dashed line) which has been corrected for. Note that we limit our results to $k<10h\mathrm{Mpc}^{-1}$ here, since our main aim is to create a reliable emulator which can be applied to observational data. The noise level would allow us to study the power spectra at even smaller scales, but the simulations may not have a high-enough force resolution in that regime.

The second column of Fig.~\ref{fig:power} shows the binned response $B(k)$ described above. As one can see from the amplitude of $B(k)$, \fr leads to an enhancement in the power spectrum on top of the difference between the theoretical prediction and the simulations. This enhancement is larger for stronger \fr models, and it depends on the $k$-scale as expected from previous works \citep{li2013,winther2015,arnold2019,arnold2018}.

\subsection{Power spectrum smoothing}
\begin{figure*}
	\includegraphics[width=\textwidth]{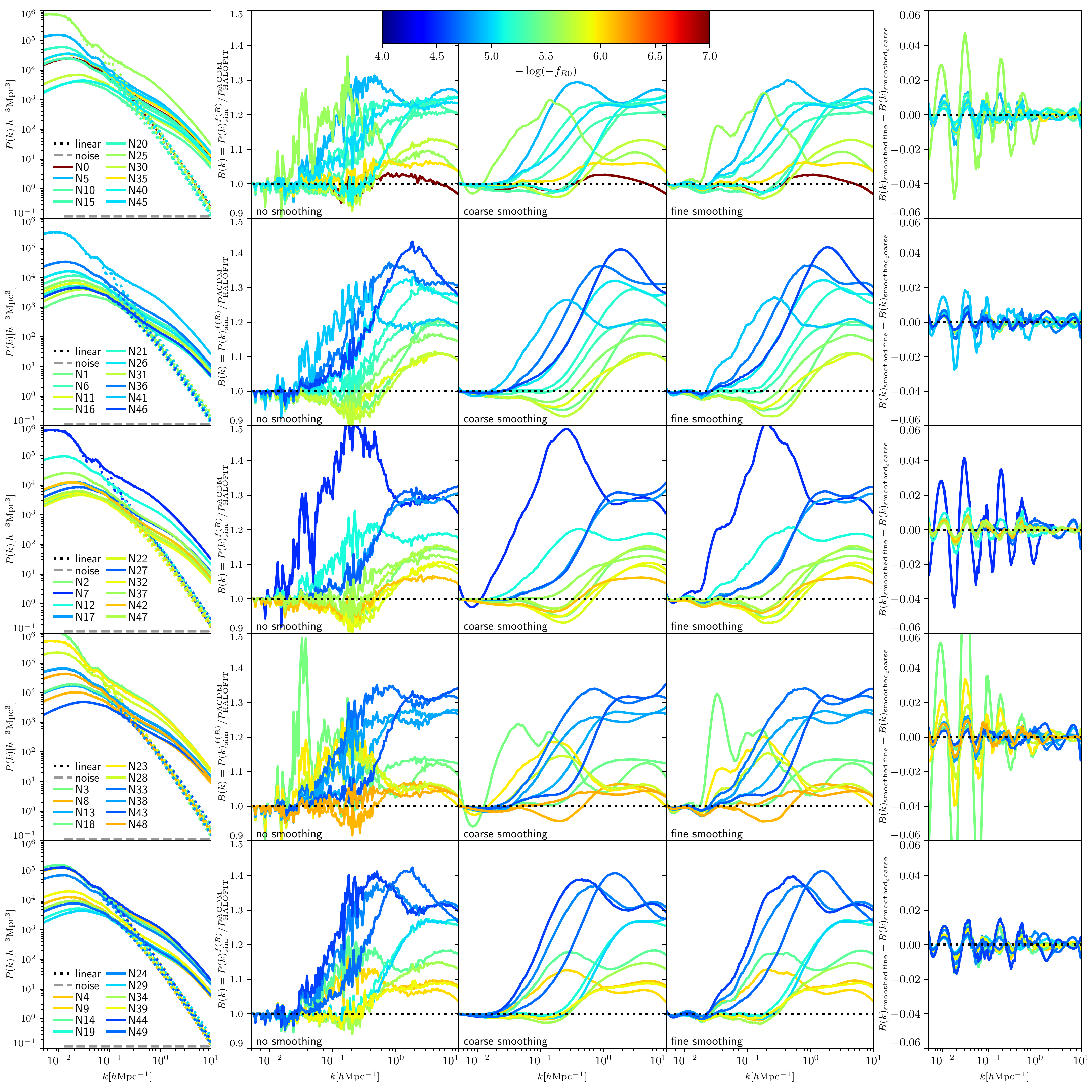}
    \caption{(Colour Online) The 3D matter power spectra for the 50 nodes (N0-N49) of cosmological parameters as measured from the 
    \fr simulations at $z=0$. To avoid over-crowded plots, the node data has been distributed over the 5 rows of the figure. The left hand panels show the binned power spectra (averaged over all box-sizes and implementations, shot noise corrected; see text for details) from the simulations as solid lines. The line colours indicate the background field of the $f(R)$ model for each node as indicated by the colour-bar at the top of the figure. Node 0 is \lcdm, but shown as $-\log_{10}\left(-\bar{f}_\mathrm{R0}\right) = 7$. The corresponding linearly evolved initial power spectra are shown as dotted lines in the same colours in the left panels, where the simulation shot-noise level is indicated by the grey dashed lines. The second column shows the binned ratio $B(k)$ between the simulated \fr power spectrum and the \textsc{halofit} \lcdm prediction for the cosmology. The third and fourth columns show the same quantity but with a coarse and fine Savitzky-Golay smoothing applied (see text), respectively. The fifth column shows the difference between the fine and the coarse smoothed $B(k)$. Black dotted lines indicate the \textsc{halofit} predictions.}
    \label{fig:power}
\end{figure*}

\begin{figure*}
	\includegraphics[width=\textwidth]{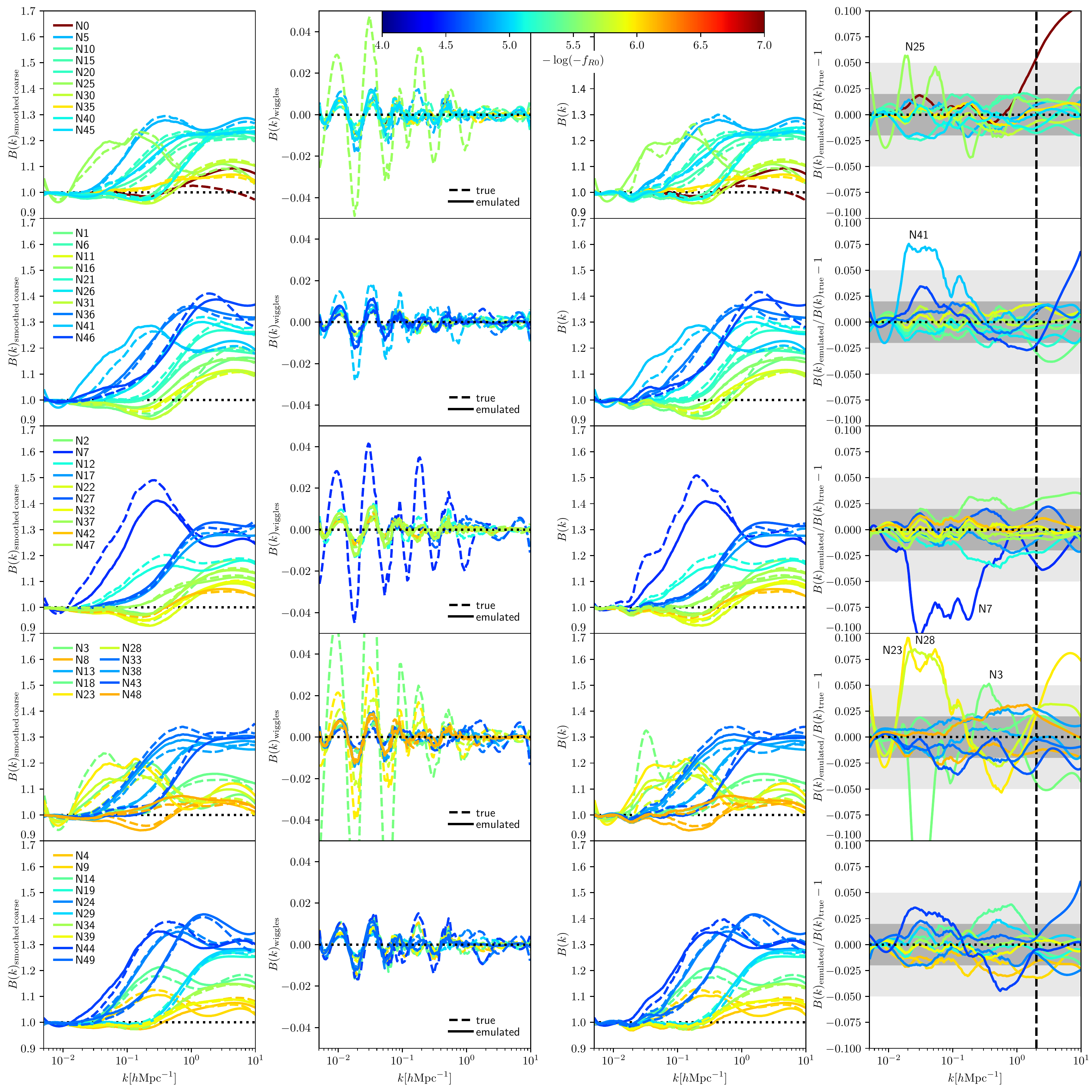}
    \caption{(Colour Online) Cross validation for the matter power spectrum emulator at $z=0$. The power spectrum for each nodes is predicted by training the emulator on the 49 other nodes. The results are split into the 5 different rows of the plot for better readability. Dashed lines show the true data for each node, and solid lines the emulator predictions from the cross validation. The colours indicate the strength of the $f(R)$ model as shown by the colour-bar at the top of the plot. The left column shows the coarsely smoothed $B(k)$. The center-left column shows the separately emulated BAO `wiggle' enhancement. The center right column shows emulated and true total $B(k)$ (i.e., the sum of the left two left hand columns). The right column shows the relative difference between the emulated and the true $B(k)$. The dark and light gray shaded regions in the right panels indicate the $\pm2\%$ and $\pm5\%$ error margins, respectively. Black dotted lines indicate equality, the dashed black vertical lines are at $k = 2\, h\mathrm{Mpc}^{-1}$.}
    \label{fig:power_emulated}
\end{figure*}

\begin{figure*}
	\includegraphics[width=\textwidth]{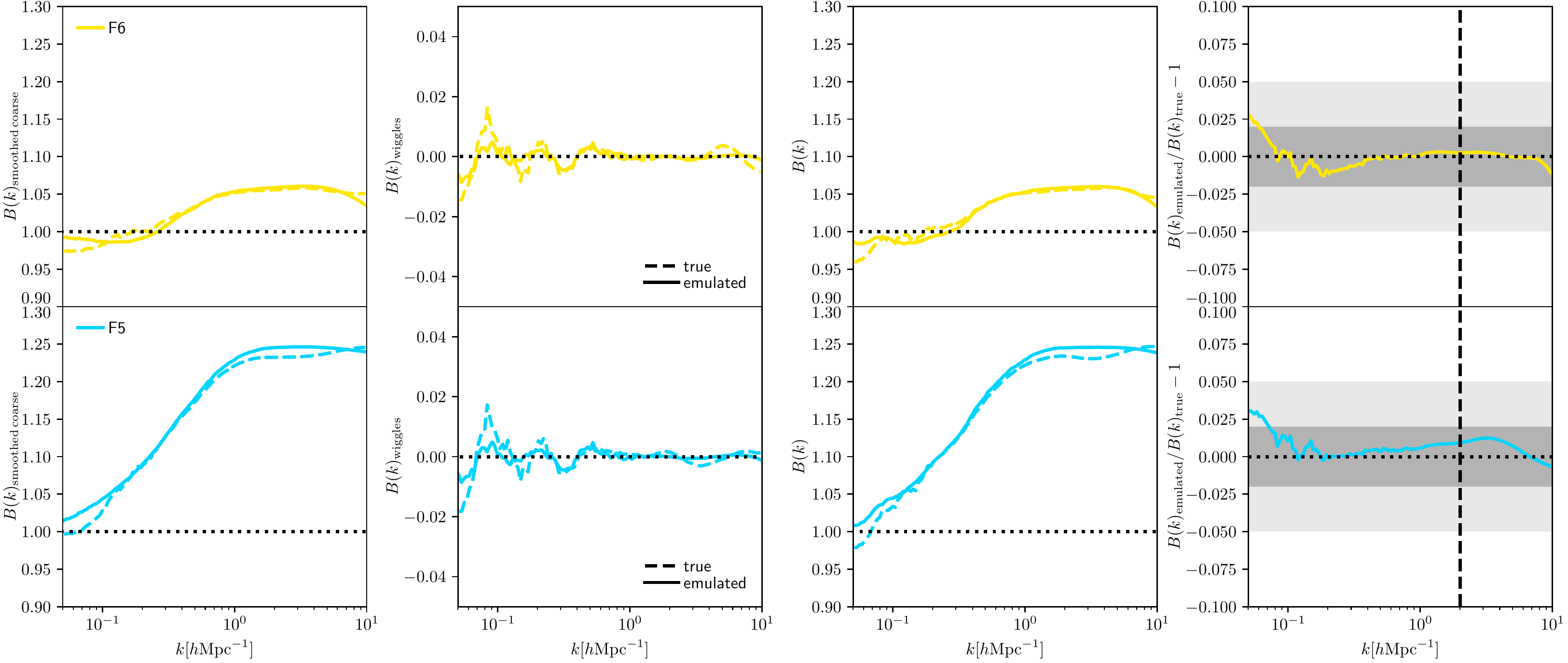}
    \caption{(Colour Online) The power spectra responses for two additional cosmologies using fiducial base parameters (like node 0) but $|f_\mathrm{R0}| = 10^{-6}$ (top) and $|f_\mathrm{R0}| = 10^{-5}$ (bottom), at $z=0$. The dashed lines show the measured spectra from the sets of 8 simulations per parameter set, while the solid lines show the emulator predictions. As in Figure \ref{fig:power_emulated}, the left column shows the coarsely smoothed response, $B(k)$, the center left column the BAO wiggle enhancement, the center right column the total $B(k)$, and the right column the relative error of the emulated $B(k)$. The dotted horizontal lines indicate equality, and the dashed vertical line in the right column shows $k=2\hompc$. The dark and light grey regions in the right column indicate $\pm 2\%$ and $\pm 5\%$ error intervals, respectively.}
    \label{fig:power_prediction}
\end{figure*}

Gaussian process emulators generally work best for $\mathcal{O}(1)$ quantities with as few minima and maxima as possible. As the power spectrum itself spans several orders of magnitude, we choose to emulate the ratio $B(k)$ to the \textsc{halofit} \lcdm prediction. As one can see from column 2 of Figure \ref{fig:power}, the (finely) binned $B(k)$ is very noisy, which poses a problem for the emulator. This problem can be overcome by applying a Savitzky-Golay filter \citep{savgolfilter} to $B(k)$, following \citet{Ramachandra2020}. {A similar technique has been employed in \citet{Euclid_emulator:2018mlb} and \citet{giblin2019}.} The filter fits a certain number, $N$, of data points left and right of the point of interest with a polynomial of $p$-th order. We have applied this technique to $B(k)$ using $N=51$ and $p=3$ to obtain the coarsely smoothed $B(k)$ in column 3 of Figure \ref{fig:power}. 

This data is now better suited for a Gaussian process emulator, but lacks the enhancement of the BAO fluctuations due to \fr because of the aggressive smoothing. To capture these enhancements as well, we perform a second, finer, smoothing of the initially binned $B(k)$ but use $N=41$ and $p=4$ to obtain the curves in column 4 of the figure. The data in column 4 is less suited for a GP emulator due to large number of minima and maxima on top of the enhancement. We therefore choose to emulate the BAO `wiggles' separately after isolating them by taking the difference between the fine and the coarse smoothed $B(k)$ (see column 5 of Figure \ref{fig:power}). 

\subsection{Emulating the 3D matter power spectrum}

We emulate separately the coarsely smoothed $B(k)$ and the BAO-induced `wiggles' described above employing the Gaussian process emulator similar to that described in \citet{slics}. 
This emulator uses the publicly-available {\sc scikit-learn} Gaussian process regression code\footnote{\url{https://scikit-learn.org/stable/modules/gaussian_process.html}}. Following numerous previous works \citep[for example,][]{habib2007,Heitmann2009,kwan2015}, we adopt a radial basis function to model the covariance of the power spectrum response in the Gaussian process. This functional form for the covariance is well motivated, assuming only that the emulated statistic varies smoothly with changes in the cosmological parameters, and that it converges 
towards the training set measurements at the simulation nodes. Training the emulator corresponds to fitting the five free hyperparameters of this model, one amplitude and four correlation lengths (one per dimension in our cosmological parameter space), by applying a gradient ascent optimisation algorithm to a likelihood distribution conditioned on the simulated training data. We refer the interested reader to \citet{gaussian_process} for more details on emulation with Gaussian processes.

On both the coarsely smoothed $B(k)$ and BAO-induced `wiggles', we perform a principal component analysis (PCA) using 4 basis functions prior to training, finding this compression of the data beneficial for improving emulation accuracy and reducing training time. We confirmed that this number of basis functions is enough to reliably emulate both the wiggles and the smoothed $B(k)$ and that a larger number does not improve the emulator performance further. 

Since the emulator {cannot model the \lcdm response} with a value of $|\bar{f}_\mathrm{R0}| = 0$ for node 0, we set this parameter to a value where no deviation of the \fr power spectrum from its \lcdm counterpart is expected. For $z=0$, this is $|\bar{f}_\mathrm{R0}| = 10^{-7.5}$, at higher redshift we use a slightly larger value to ensure a smooth transition.  

{The final power spectrum prediction $P_\mathrm{prediction}(k)$ is obtained from the emulated coarsely smoothed $B_{\mathrm{coarse}}(k)$ and the emulated `wiggles' $W(k)$:
\begin{align}
    P_\mathrm{emulated}(k) = P_\mathrm{HALOFIT}^{\Lambda \mathrm{CDM}} \times \left[B_{\mathrm{coarse}}(k) + W(k) \right].
\end{align}}

The training described above is repeated at all common output redshifts available. In the following section we verify the emulator predictions for $z=0$, while higher-redshift results are presented in Appendix \ref{emu_ver}.

\subsection{Emulator validation}
\label{emu_val}

The first step we take towards validating our emulator predictions is performing a `leave-one-out' cross validation test. This method removes one node from the sample a time, trains the emulator on the remaining 49 nodes and then predicts the power spectrum response $B_{\rm coarse}(k)$ and the wiggle enhancement $W(k)$ due to $f(R)$ gravity at the excluded node. This process is repeated for all nodes,  with results shown in Figure \ref{fig:power_emulated}, where the dashed lines show the true data for each node, and solid lines the predictions of the emulator. The two separately emulated quantities, the coarsely smoothed $B(k)$ and the wiggles, are shown in the left two columns of the plot. 

As one can see, the emulator performs well except for some of the more extreme cosmologies. This is also reflected in the accuracy of the {predicted} total $B(k)$ shown in the right two columns, where the rightmost column is the relative difference between the emulated and the true simulated $B(k)$. For most nodes, the latter remains entirely within the $\pm 5\%$ error margin indicated by the light grey bands in the rightmost column of Fig.~\ref{fig:power_emulated} for $k<2\hompc$. The nodes with a $>5\%$ error for $k<2\hompc$ are 25, 41, 7, 23, 28 and 3. These nodes are highlighted in red in Fig.~\ref{fig:param}, from which it is obvious that all of them feature an extremely low $\Omega_\mathrm{m}$ value, resulting in a large baryon fraction. For example, for node 3, approximately $50\%$ of the matter is represented by baryons, leading to much larger than normal BAO wiggles that are further enhanced during cosmic structure formation. The emulator struggles to emulate these very large wiggles for node 3 (see column 2, row 4 of Fig.~\ref{fig:power_emulated}), resulting in a poor prediction accuracy at large scales in that specific case. Similar effects of the large baryon fraction can be observed for the other low-$\Omega_{\rm m}$ nodes. We note that values of $\Omega_\mathrm{m} \approx 0.1$ are far from the current observational constraints $\Omega_\mathrm{m} =  0.321 \pm 0.013$ \citep[][TT+lowE]{planck2018} and the poorer performance of our emulator for these nodes will therefore not affect its applicability in practice. {Although this subset of nodes may have extreme baryon fractions, it is necessary to include them in our ensemble of cosmologies to cover the range of $\Omega_{\rm m}$ values consistent with the constraints from current weak lensing surveys, whilst keeping $\Omega_{\rm b}$ fixed to minimise the dimensionality of our emulation parameter space.} 
It is also apparent from Figure \ref{fig:power_emulated} that the emulated power spectra for the majority of the nodes have an error smaller than $2\%$ for $k<10\hompc$, as shown by the dark grey bands in the rightmost column. 

As a comparison, we have trained the emulator directly on the \textsc{halofit} nonlinear $P(k)$ for the same 50 cosmologies (i.e., \fz{} is set to $0$) and found a similar emulator accuracy for the same cross validation exercise. This means that the accuracy is limited by the number of nodes in general, and is not driven by the modified gravity sector.

As a further step towards the verification of our emulator, we predicted the matter power spectra for two cosmologies which were not part of our training, and compared the emulator predictions to two sets of $f(R)$ simulations, each consisting of 8 independent runs with $1024^3$ particles in $500\mpcoh$ boxes {(the ICs of these 8 runs are paired in the same way as for the ones used to train the emulator, to suppress sample variance on large scales)}. This is thus not only a test of how well the emulator can predict the matter power spectra for different cosmologies, but also a measure of how much cosmic variance affects the results, as the 8 simulations per parameter set offer a significantly larger statistical sample. Both simulation sets use the same cosmological base parameters as node 0, but a non-zero modified gravity parameter, \fz, respectively $-10^{-6}$ (F6) and $-10^{-5}$ (F5). The results are shown in Figure \ref{fig:power_prediction}, where the columns are the same as in Figure \ref{fig:power_emulated}. 
We do not have large box implementations for these cosmologies, 
and therefore limit the $k$-range of the simulation power spectra to $k>5\times10^{-2}\hompc$. As one can see from Figure \ref{fig:power_prediction}, the emulator works very well over the whole $k$-range shown, with a maximum relative error of roughly $2.5\%$ for F6 and $3\%$ for F5. The relative errors for $0.07<k/(\hompc)<10$ stay completely within the $2\%$ error margin for both parameter-sets, showing that our emulator can robustly predict the power spectrum response relative to \textsc{halofit}. {The high accuracy achieved at such nonlinear scales are critical for weak lensing science, which is mostly sensitive to these physical scales.} We plan to use these $f(R)$ simulations in upcoming cosmic shear data analyses.

\section{Conclusions}
\label{sect:con}

We presented a new suite of 200 cosmological simulations in \fr which explore a wide  cosmological parameter space for 50 different combinations of $\Omega_{\rm m}$, $h$, $\sigma_8$ and $\fz$. 
These simulations are designed to aid the building of emulators for weak lensing, the halo mass function (which will both be presented in future works), and a matter power spectrum emulator which we present in this work. 


The simulations combine two different resolutions per parameter set, allowing to probe a wide range of cosmological scales and dynamical range. For each resolution, two implementations with independent initial conditions are performed. The ICs for these are selected such that the large-scale cosmic variance errors of the two implementations approximately cancel each other. We then combine the data from the four individual runs per cosmological parameter set using a fine binning technique. The noise is subsequently removed from the combined power spectrum using a Sawitzki-Golay filter. In order to provide a smooth order-unity quantity as our training data, we choose to emulate the ratio of the simulation power spectrum to the \textsc{halofit} \lcdm predicted power spectrum, in other word  a proxy for the relative difference between \fr and GR, using an easily accessible GR baseline. We also choose to emulate the enhancement of the BAO wiggles separately as this gives a better overall performance. The emulator is run at different redshifts between $z=0$ and $z=2$ independently and is made publicly available online. 

We have tested the performance of the power spectrum emulator using cross validation, finding that it has an accuracy better than $5\%$ for all nodes with $\Omega_\mathrm{m} > 0.18$ at all emulated redshifts. Nodes with even lower $\Omega_\mathrm{m}$ have a large baryonic-to-total matter fraction, as $\Omega_\mathrm{b}$ remains unchanged, which leads to strong BAO 
wiggles which our emulator struggles to predict. This will nevertheless not limit the practical applicability of our emulator, as it performs very well (better than $5\%$) for reasonable choices of $\Omega_\mathrm{m}$. We have also tested the emulator predictions against simulations for parameter sets which are not part of the training set (fiducial cosmology but
with different values of $\fz$), and found that the emulator can predict those to better than $2\%$ accuracy. As an additional check, we have also trained the emulator using nonlinear matter power spectra for $50$ $\Lambda$CDM models with the same cosmological parameters (but $\fR=0$), generated by \textsc{halofit}, and obtained the same accuracy in the emulated $P(k)$. This implies that the $5\%$ accuracy in the cross-validation test is not due to the noisy simulation data and our smoothing of them, or the effect of $f(R)$ enhancement, but is more likely caused by the small size of the training dataset, especially considering that our $50$ nodes covers very wide parameter ranges. The $5\%$ maximum error should be considered as the worst-case scenario, not only since the other test suggests a $2\%$ emulator accuracy, but also because even in the cross-validation case a $2\%$ accuracy is achieved for the majority of nodes at $k<10h\mathrm{Mpc}^{-1}$

We conclude that our simulation suite is suitable for building emulators for different cosmological observables in \fr. These will be presented in subsequent works. The power spectrum emulator, in particular, can be used to make quick and accurate predictions for the fully nonlinear matter power spectrum in \fr, e.g., to analyse observational data using an MCMC pipeline over a wide $k$-range. This emulator does not include baryonic effects, but as shown by \citet{arnold2019}, baryonic and \fr effects on the matter power spectrum can be estimated, to a good precision for reasonable $f(R)$ models, from separate baryonic \lcdm simulations and \fr dark matter-only simulations. 
This opens the possibility to combine our emulator with a baryonic physics power spectrum emulator for future analysis. Finally, in this paper we have focused on a particular $f(R)$ model, and in subsequent works we will run and present simulations and emulators for other modified gravity models.

\section*{Acknowledgements}

The authors thank Ben Bose for useful discussions and for providing matter power spectrum predictions generated with the ReACT code. 
{CA and BL are supported by the European Research Council (ERC) through a starting Grant (ERC-StG-716532 PUNCA). BL is further supported by the UK Science and Technology Funding Council (STFC) Consolidated Grant No.~ST/I00162X/1 and ST/P000541/1.} 
{JHD acknowledges support from an STFC Ernest Rutherford Fellowship (project reference ST/S004858/1)}
{BG acknowledges the support of the Royal Society through an Enhancement Award (RGF/EA/181006).} 
{YC acknowledges the support of the Royal Society through a University Research Fellowship and an Enhancement Award.}
{This work used the DiRAC@Durham facility managed by the Institute for Computational Cosmology on behalf of the STFC DiRAC HPC Facility (\url{www.dirac.ac.uk}). The equipment was funded by BEIS via STFC capital grants ST/K00042X/1, ST/P002293/1, ST/R002371/1 and ST/S002502/1, Durham University and STFC operation grant ST/R000832/1. DiRAC is part of the UK National e-Infrastructure.}

\section*{Data Availability}

The matter power spectrum training data, as well as the power spectrum emulator, are publicly available from the following git repository: \url{https://bitbucket.org/arnoldcn/forge\_emulator/}. The simulation outputs can be made available upon reasonable request to the authors. 

\appendix

\section{Emulator performance at higher redshifts}
\label{emu_ver}

As mentioned in the main part of the paper, the 3D power spectrum emulator is available not only at $z=0$ but also for higher redshifts. In this Appendix, we show the cross-verification results for $z=1$ and $z=2$. At even higher redshifts, the \fr effect on the power spectrum is expected to be negligible on the scales predicted by our emulator. The ratio between the power spectrum measured from our simulations and \textsc{halofit} is therefore dominated by the inaccuracies of both and cannot reliably be predicted by an emulator. Fortunately though, the need for an emulator at those higher redshifts is not as strong as for low $z$ anyway.

\begin{figure*}
	\includegraphics[width=\textwidth]{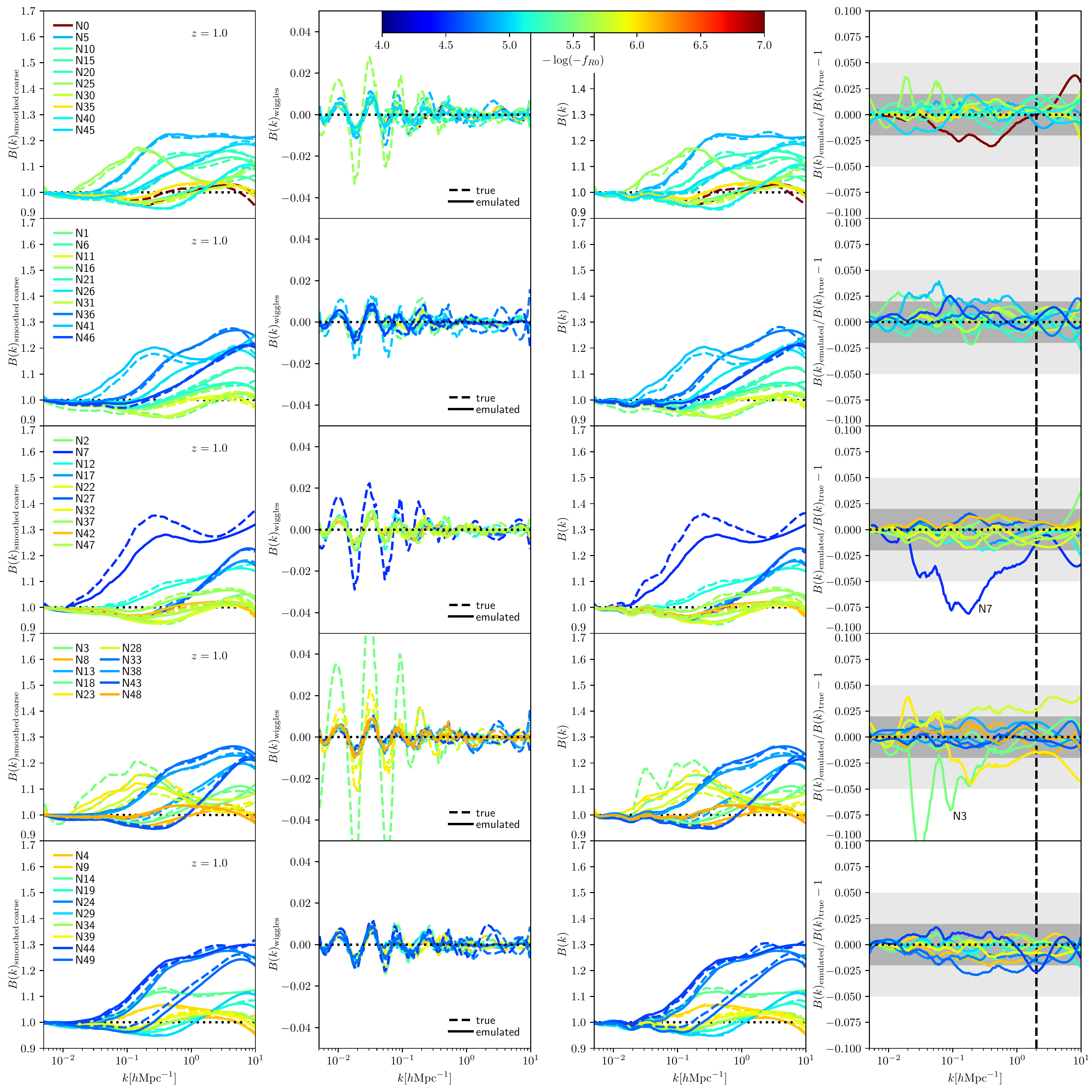}
    \caption{(Colour Online) Same as Figure \ref{fig:power_emulated} but at $z=1$.}
    \label{fig:power_emulated_z1}
\end{figure*}

\begin{figure*}
	\includegraphics[width=\textwidth]{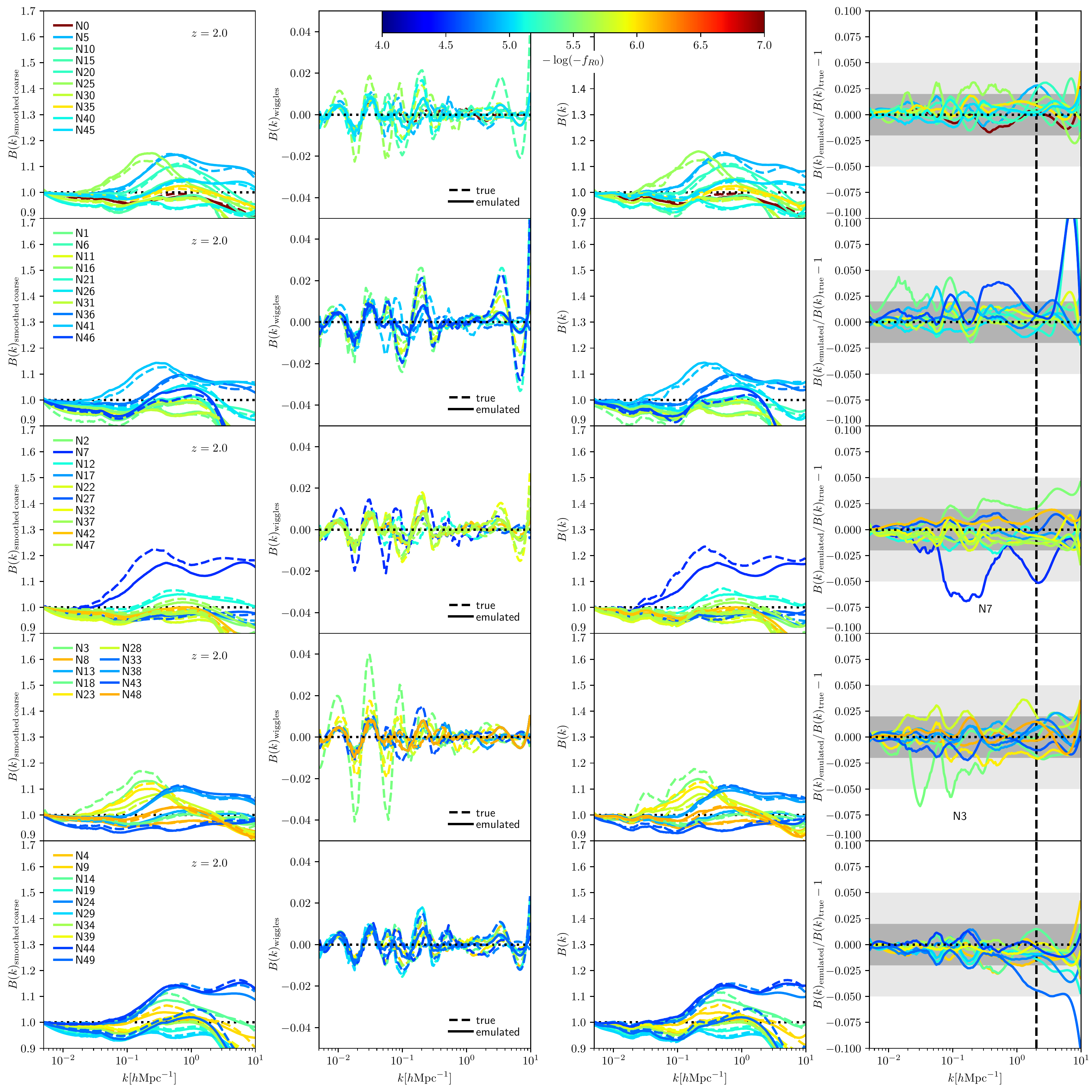}
    \caption{(Colour Online) Same as Figure \ref{fig:power_emulated} but at $z=2$.}
    \label{fig:power_emulated_z2}
\end{figure*}

As one can see from Figures \ref{fig:power_emulated_z1} and \ref{fig:power_emulated_z2} the emulator produces very accurate results for both redshifts. At $z=1$ all nodes except nodes 3 and 7 --- which have a very low value of $\Omega_\mathrm{m}$ (see discussion for $z=0$ in Section \ref{emu_val}) --- are within the $5\%$ error margin and most nodes are completely within the $2\%$ error band. For $z=2$ we find a similar accuracy from the cross-validation tests.

To keep this section brief, we do not show the results for the intermediate redshifts for which the emulator is also available here. We did nevertheless check that the emulator performance is better than at $z=0$ for all available redshifts. 

\section{Comparison with other $f(R)$ matter power spectrum predictions}

\begin{figure}
	\includegraphics[width=\columnwidth]{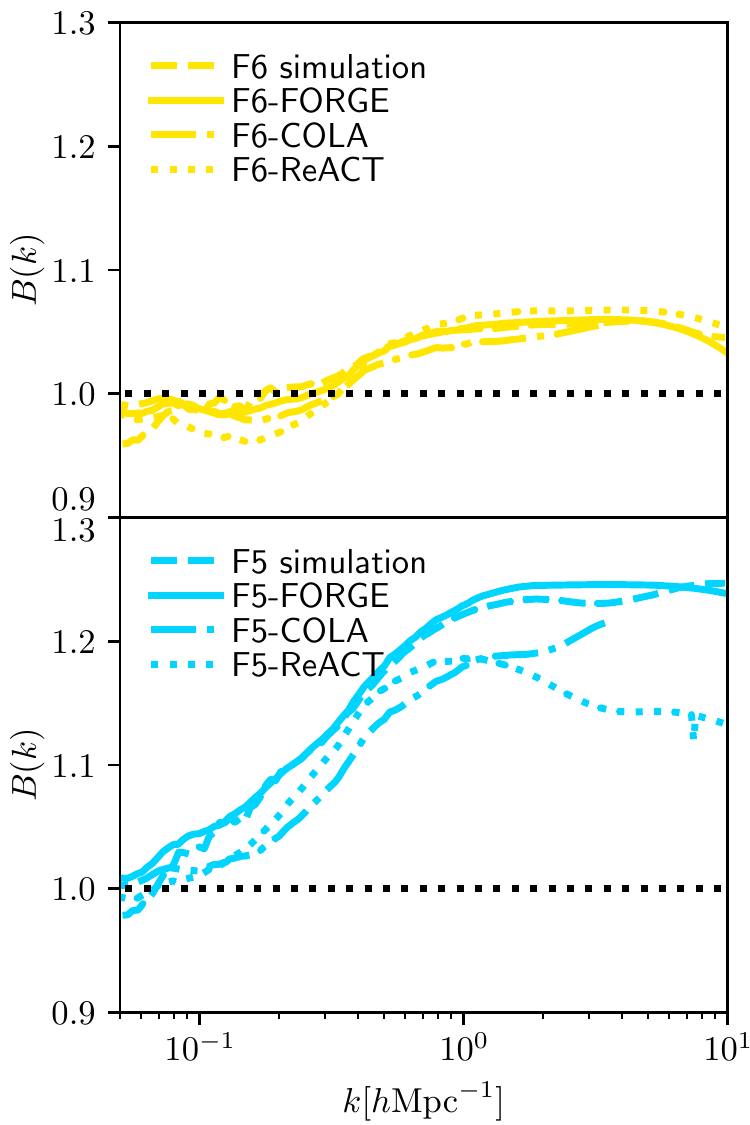}
    \caption{(Colour Online) The average power spectrum response $B(k)$ measured from the 8 node-0 cosmology simulations (dashed lines) for F6 (top panel, yellow) and F5 (bottom panel, cyan) compared to several different predictions. The response from the \textsc{FORGE} emulator presented in this paper are shown as solid lines, predictions made with the emulator based on \textsc{mg-cola} simulations \protect\citep{Ramachandra2020} as dash-dotted lines, and the $B(k)$ given by \textsc{ReACT} \protect\citep{reactA, reactB, reactC} 
    as dotted lines. Black dotted lines indicate unity. The \textsc{ReACT} and \textsc{MG-COLA} predictions are corrected for the difference between \lcdm simulations and \textsc{halofit} in this plot (see the text for more details).}
    \label{fig:power_compare}
\end{figure}

To compare the performance of the \textsc{FORGE} power spectrum emulator with other 
theoretical predictions for the matter power spectrum in \fr, we plot the predictions of different 
approaches for $B(k)$, as well as our simulation results for F5 and F6 and node-0 background cosmology, in Figure \ref{fig:power_compare}. As we have already seen from Figure \ref{fig:power_prediction}, the \textsc{FORGE} emulator matches the result of the 8 F6 and F5 simulations (reproduced here as dashed lines) on a $1-2\%$ level for scales up to $k=10\hompc$. Further, we have shown the predictions by an $f(R)$ power spectrum emulator constructed using \textsc{mg-cola} simulations\footnote{\url{https://github.com/LSSTDESC/mgemu.}} \citep[][ dash-dotted lines]{Ramachandra2020}, and by the reaction method \textsc{ReACT}\footnote{\url{https://github.com/nebblu/ReACT/tree/react_with_neutrinos}.} \citep[][coloured dotted lines]{reactA, reactB, reactC}.
As the two additional approaches give $P(k)$ or its enhancement with respect to $\Lambda$CDM, rather than $B(k)$, to get the latter we used the definition, Eq.~\eqref{eq:Bk_def}, with the numerator replaced by the predictions of these approaches.

For F6, the \textsc{mg-cola} 
emulator prediction agrees equally well with the simulations within its range of validity (the authors do not recommend using it for scales of $k > 1 \hompc$), but it differs from the simulation result by about $5$--$7\%$ for F5. 
\textsc{ReACT} also agrees very well with the simulation result for F6, while for F5 it agrees within about $5\%$ for $k<0.7\hompc$ but then starts to deviate significantly on smaller scales. The difference between \textsc{ReACT} and our F6/F5 simulations (and therefore \textsc{FORGE}) is likely because in the theoretical prediction by the former the fitting functions for $\Lambda$CDM (as opposed to $f(R)$ model) halo mass functions and concentration-mass relations are used (these will be completed in upcoming public versions of \textsc{ReACT}). 

To summarise the performance of different power spectrum predictions, one can say that all these methods perform well for F6, but there are differences for stronger $f(R)$ models. The good agreement in F6 is not surprising, since the power spectrum in this model differs from that in $\Lambda$CDM by at most $\simeq5\%$ for the whole $k$ range. While the \textsc{FORGE} emulator 
reproduces the simulation result for F5 to great accuracy as well, the other two methods show stronger deviations. This result has to be taken with caution though, for the reason mentioned above, and also because \textsc{ReACT} and \textsc{mg-cola} have been calibrated or trained with simulations produced with other simulation codes, which might lead to up to a few percent differences in the training data already.

\section{How to use the power spectrum emulator}
\label{emu_user_guide}

The \textsc{FORGE} matter power spectrum emulator and the power spectrum response data are publicly available from a git repository at \url{https://bitbucket.org/arnoldcn/forge\_emulator/}. The repository contains all the training data used to train the emulator, as well as the emulator package in the form of a \textsc{python} module (\texttt{GPR\_Emulator.py}). This module can be imported into any \textsc{python} application. The repository also contains an example for a cross-validation application of the emulator using the power spectrum data, as well as an example script for training the emulator on the power spectrum data and making a prediction for a certain parameter combination. Note that once the emulator has been trained on the power spectrum data once, its internal state is saved and will be re-loaded for predictions. It is hence not necessary to re-train the emulator before making predictions enabling MCMC applications. A more detailed breakdown of the package functionalities is given in a README included in the repository.

\bibliographystyle{mnras}
\bibliography{paper} 

\bsp
\end{document}